\documentclass[twosides]{IEEEtran}

\usepackage{cite}
\usepackage{amsfonts,amsmath,amssymb}
\usepackage{graphicx,color,epsfig,rotating}
\usepackage{latexsym}
\usepackage{subfigure}
\usepackage{epic,eepic}
\usepackage{algorithm,algorithmicx,algpseudocode}
\usepackage{float}
\usepackage{multirow}
\usepackage{bbm}
\usepackage{bm}
\usepackage{url}
\usepackage{lineno}
\usepackage{enumerate,enumitem}
\usepackage{amsthm}
\usepackage{verbatim}
\usepackage{setspace}

\floatname{algorithm}{Algorithm}

\algnewcommand\algorithmicswitch{\textbf{switch}}
\algnewcommand\algorithmiccase{\textbf{case}}
\algnewcommand\algorithmicassert{\texttt{assert}}
\algnewcommand\Assert[1]{\State \algorithmicassert(#1)}%

\algdef{SE}[SWITCH]{Switch}{EndSwitch}[1]{\algorithmicswitch\ #1\ \algorithmicdo}{\algorithmicend\ \algorithmicswitch}%
\algdef{SE}[CASE]{Case}{EndCase}[1]{\algorithmiccase\ #1}{\algorithmicend\ \algorithmiccase}%
\algtext*{EndSwitch}%
\algtext*{EndCase}%

\begin{document}

\title{{\Huge Tidal-Like Concept Drift in RIS-Covered Buildings: When Programmable Wireless Environments\\ Meet Human Behaviors} }
\author{ \IEEEauthorblockN{Zi-Yang Wu,~\emph{Member, IEEE}, Muhammad Ismail,~\emph{Senior Member, IEEE},\\ Jiliang Zhang,~\emph{Senior Member, IEEE}, and Jie Zhang,~\emph{Senior Member, IEEE}}

\thanks{Date of current version May 19, 2025. The work of Z.-Y. Wu was supported in part by China Postdoctoral Science Foundation under Award 2023T160087, and Liaoning Provincial Natural Science Foundation Program under Grant 2023-MSBA-076. The work of M. Ismail is supported by the National Science Foundation (NSF) Award 2138234. Accepted from open call. (\textit{Editor: Kun Yang})
(\textit{Corresponding author: Jiliang Zhang}.) 

Z.-Y. Wu is with The State Key Laboratory of Synthetical Automation for Process Industries and The College of Information Science and Engineering, Northeastern University, Shenyang, China. (email: wuziyang@neu.edu.cn)

M. Ismail is with Cybersecurity Education, Research, and Outreach Center (CEROC) and Department of Computer Science, Tennessee Tech University, TN, USA. (email: mismail@tntech.edu)

J. Zhang is with The State Key Laboratory of Synthetical Automation for Process Industries and The College of Information Science and Engineering, Northeastern University, Shenyang, China, and also with The National Mobile Communications Research Laboratory, Southeast University, China. (email: zhangjiliang1@mail.neu.edu.cn)

Jie Zhang is with the R\&D Department, Cambridge AI+ Ltd., CB23 3UY Cambridge, U.K., and also with the R\&D Department, Ranplan Wireless Network Design Ltd., CB23 3UY Cambridge, U.K. (email: Jie.Zhang@ranplanwireless.com)
}}

\markboth{Preprint edition, accepted by IEEE Wireless Communications, May 2025}{WU \MakeLowercase{\textit{et. al.}}: Concept drift in RIS-covered Buildings}

\maketitle

\begin{abstract}
Indoor mobile networks handle the majority of data traffic, with their performance limited by building materials and structures. However, building designs have historically not prioritized wireless performance. Prior to the advent of reconfigurable intelligent surfaces (RIS), the industry passively adapted to wireless propagation challenges within buildings. Inspired by RIS's successes in outdoor networks, we propose embedding RIS into building structures to manipulate and enhance building wireless performance comprehensively. Nonetheless, the ubiquitous mobility of users introduces complex dynamics to the channels of RIS-covered buildings. A deep understanding of indoor human behavior patterns is essential for achieving wireless-friendly building design. This article is the first to systematically examine the tidal evolution phenomena emerging in the channels of RIS-covered buildings driven by complex human behaviors. We demonstrate that a universal channel model is unattainable and focus on analyzing the challenges faced by advanced deep learning-based prediction and control strategies, including high-order Markov dependencies, concept drift, and generalization issues caused by human-induced disturbances. Possible solutions for orchestrating the coexistence of RIS-covered buildings and crowd mobility are also laid out.
\end{abstract}

\begin{IEEEkeywords}
Millimeter wave, visible light communications, programmable wireless environments, reconfigurable intelligent surfaces, mobility, building wireless performance.
\end{IEEEkeywords}

\IEEEpubidadjcol
\IEEEpubid{\begin{minipage}{\textwidth}
\footnotesize
\centering
\vspace{15mm}
\copyright 2025 IEEE. Personal use is permitted, but republication/redistribution requires IEEE permission.\\ See https://www.ieee.org/publications/rights/index.html for more information.
\end{minipage}
}

\section{Introduction}
\label{sec:intro}

Reconfigurable intelligent surface (RIS) is considered a pivotal innovation supporting the $6^\text{th}$ generation and beyond ($6$G+) mobile communications, marking a milestone in actively intervening in electromagnetic propagation to optimize communication networks. For enhancing outdoor networks, RIS has demonstrated its powerful flexibility and vast application potential of meta-materials \cite{6711974,zhang2022metasurface}. By reshaping reflection beams to suppress the randomness of channels\cite{del2019optimally}, RIS enables the transmitted signals to reach shadowed areas beyond the field of view and to track fast-moving users, thereby gaining coverage and mobility \cite{9771330,10323414}.

\begin{figure}
\centering
\includegraphics[width=0.5\textwidth]{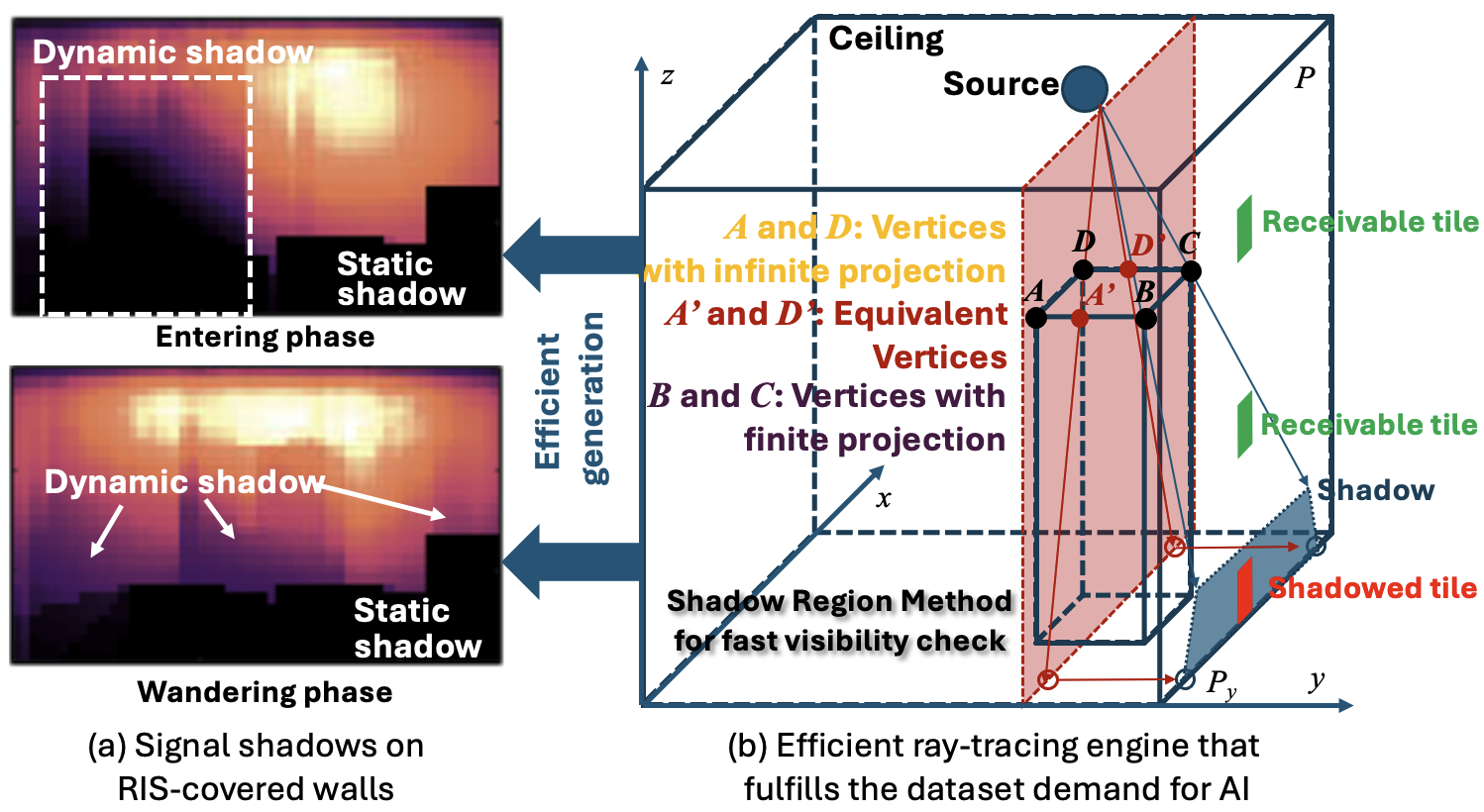}
\caption{(a) An example of the distribution of normalized channel gains on a RIS-covered wall, where the shadows represent areas with relatively low channel gains induced by moving users and indoor furnishings. The brightest region yields the highest channel gain, and the darkest one has the lowest value. (b) The proposed efficient visibility identification method accounts for infinite projections in ray-tracing, enabling the generation of a huge dataset required by learning-based methodologies \cite{wu2023efficient}.}
\label{fig:f1}
\end{figure}

However, nearly $96$\% of mobile traffic occurs indoors, a domain that even the most intelligent outdoor RIS cannot reach. The growing demand for indoor coverage and capacity necessitates denser and more efficient indoor network deployments. Simply optimizing indoor wireless networks without changing building structure cannot efficiently meet this demand. This is because building materials and layouts inherently affect signal propagation, determining the upper limits of building wireless performance (BWP) \cite{zhang2021fundamental}. The conventional network optimizations alone cannot break through these limitations. Therefore, refining the building itself becomes the key to fundamentally improving BWP. Yet, could building materials possess the ability to actively suppress channel uncertainty? This article reveals the feasibility of this idea: embedding low-cost passive RIS tiles into building materials to make RIS ubiquitous, thereby fundamentally enhancing BWP in terms of coverage and mobility\footnote{Compared to the surface-mounted mmWave RIS tiles on walls, our primary focus lies in their embedded integration within wall structures \cite{zhang2021wireless}. For visible light communications (VLC), transparent glass walls remain fully compatible with surface-mounted RIS tile implementations \cite{10323414}.}. We are making the first attempt to unfold the upper BWP limits of RIS-covered buildings constrained by indoor human behaviors.

The walls with complete RIS coverage only function under powerful control strategies. The effectiveness of these strategies depends on a deep understanding on the channel of RIS-covered buildings. User behavior patterns and the presence of objects like furniture increase the complexity of indoor environments, which conventional methods struggle to address. Meanwhile, learning-based approaches face challenges related to statistical concept drift and generalization \cite{10288574}.

This article clarifies the impact of human behavior on the channels of RIS-covered buildings to benefit artificial intelligence-based management. The effort includes re-examining the statistical representation of mobile RIS channels, where a tidal evolution pattern is discovered for the first time. Through extensive empirical statistics, we have reproduced the cross-scale interactions among macro-micro mobility and channels as snapshotted in Fig. \ref{fig:f1} driven by a novel ray-tracing engine-based efficient dataset augmentation. The users contribute to the dynamic channel shadows on the walls, which evolve following the mobility process. The macro process is governed by a bounded L\'evy process and a semi-Markov renewal process constrained by return tendencies, while the micro process is characterized by a Markov process featuring routing, turning, and terminal orientation behaviors. Although each scale—from macro-level spatio-temporal human behaviors down to micro-level RIS tile channels at each time slot—can be meticulously modeled, the combined complexity of these coupled layers disrupts statistical uniformity and instead produces tidal-like statistical drift.

The tidal evolution leads to severe concept drift. Any attempt to generalize channel statistics has exhibited a fatal deviation of more than $20$\% from practical mobility. Therefore, we assert that a generalized mobile channel model for RIS-covered buildings does not exist. Two promising $6$G+ air interfaces are elaborated, namely mmWave and visible light communications delivered by indium tin oxides (ITO) or liquid-crystal reflectors \cite{10323414}. 
We analyze the pattern evolution of the channels of RIS-covered buildings from time, space, wavelength, and user density perspectives. The challenges imposed by the concept drift through outage prediction, RIS tile exclusion, and coexistence of access points (AP) and RIS are confirmed through numerical experiments, which conclude that the planning and operation of RIS-covered buildings still call for in-depth and dedicated research efforts.

\begin{figure*}
\centering
\includegraphics[width=\textwidth]{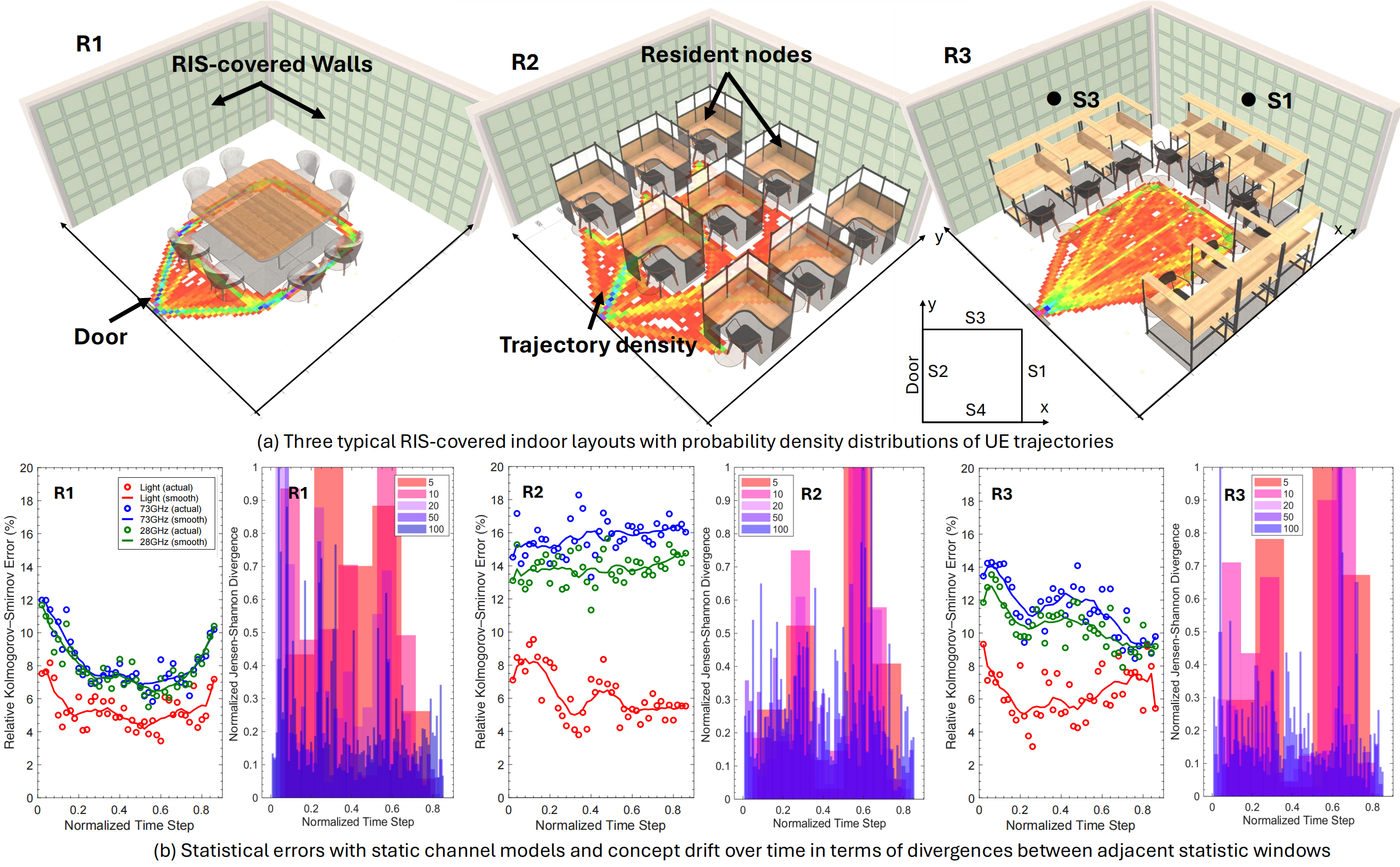}
\caption{(a) Illustration of the building layouts. Mobile users, holding UEs that access four to nine light or mmWave APs distributed evenly on the ceiling plane, are assumed in these scenarios. We introduce $1,500$ RIS tiles on each wall. Four walls in each layout are labeled with S$1$ to S$4$, where S$2$ is the wall with the door, S$1$ is located opposite to S$2$, S$3$ is located on the left side after entering the door, and S$4$ is on the right side. (b) Fitting performance over time in terms of Kolmogorov-Smirnov distances (KSD) through distribution and parameter optimizations. The parameters at each time step are optimally selected to minimize the fitting residues. The normalized Jensen-Shannon (JS) divergences between adjacent statistical windows are also illustrated, where five statistical window update frequencies are considered as $5$, $10$, $20$, $50$, and $100$.}
\label{fig:f2}
\end{figure*}

\begin{figure}
\centering
\includegraphics[width=0.5\textwidth]{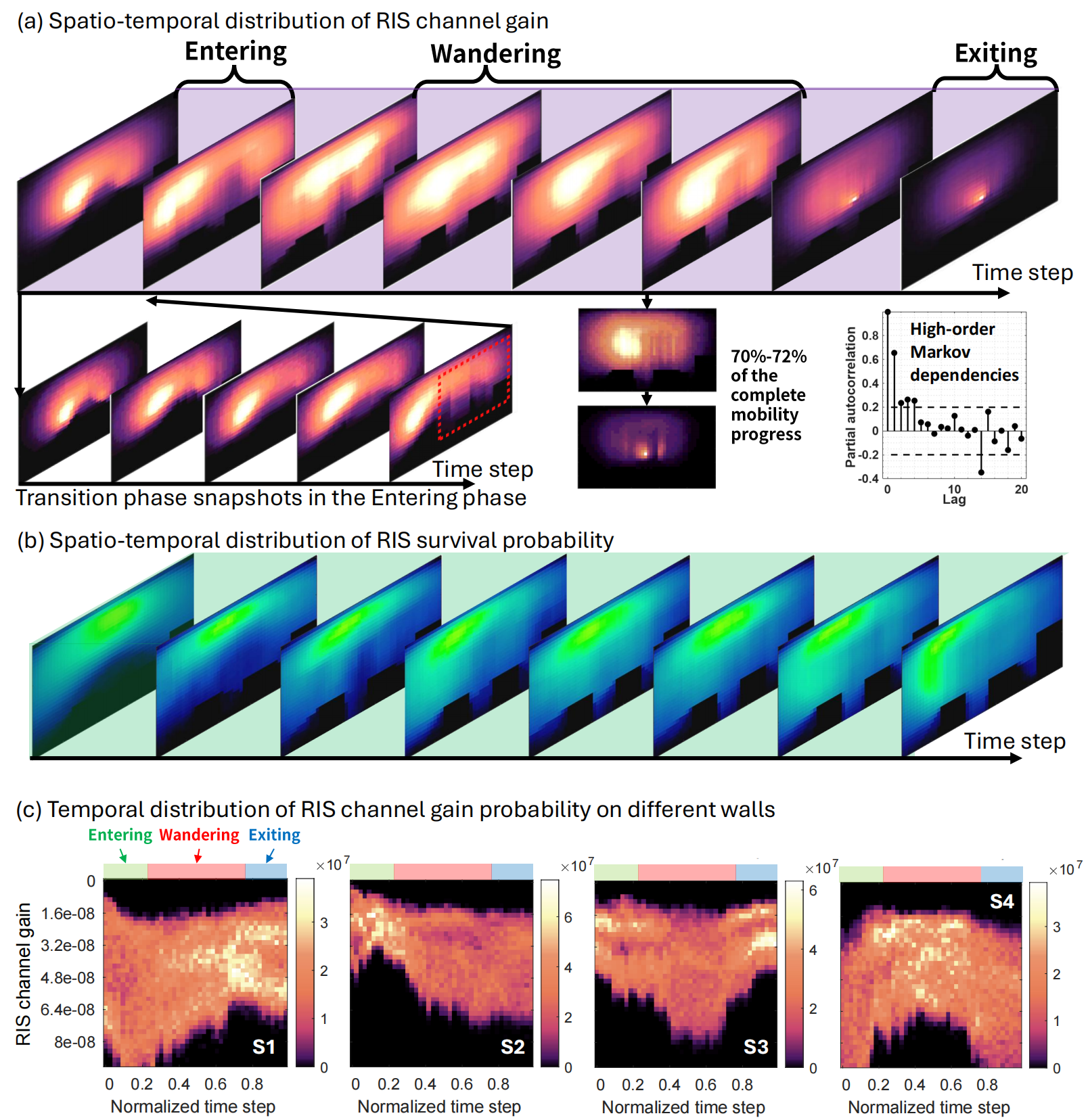} 
\caption{Spatio-temporal distribution snapshots of RIS surface channel statistics. The analysis and discussion for the presented setup can be generalized to the others. The illustrated statistics in (a) and (b) are obtained based on the S$4$ in layout R$2$ supporting eight UEs via four APs operating VL band. The distribution has been normalized in each snapshot, where the brightest region yields the highest channel gain or survival rate, and the darkest one has the lowest value. The partial auto-correlation function (PACF) of the RIS channel state is shown to demonstrate the increased Markovian order, where the PACF at a lag of 1 corresponds to the fitness of the one-order Markov process. (c) The probability density functions (PDF) of RIS channel gain and survival rate of RIS tiles over time. The instance is selected from layout R$1$ with $4$ APs and $8$ UEs under $73$ GHz mmWave support. The PDF is obtained at each time interval and the colors correspond to the probability density values.}
\label{fig:f3}
\end{figure}

\section{Inaccuracy of Conventional Generalized Statistical Channel Model}

Existing research concludes general statistical channel models under statistical experimental setups \cite{CuiTieJun2021}. Here, we step further, showing that the involvement of multi-user behavior distorts these statistical channel models, and thus hinders the robustness of conventional optimizations based on these models.

First, we introduce practical crowd behaviors into RIS-covered buildings in advance of seeking a generalized channel statistic. We take three typical layouts R$1$, R$2$, and R$3$ shown in Fig.~\ref{fig:f2}(a) as examples, meanwhile three high-frequency bands are considered, namely, $28$ GHz mmWave ($5$G Band n$261$), $73$ GHz mmWave (E-Band), and visible light (VL) bands \cite{10517366}. The mmWave and VL are at the far ends of the high-frequency bands, respectively, offering quite different diffraction capabilities. $1,500$ small RIS tiles are evenly installed on and completely cover each wall in the buildings. Each RIS tile functions independently and is designed in the size of $100$mm $\times$ $100$mm with $100$ meta-surfaces sized in $10$mm $\times$ $10$mm, such that the RIS broadcast channel in mmWave \cite{CuiTieJun2021} or VL bands should be seen as far-field. To reproduce indoor crowd behavior accurately, we set experiments and measurements on real-world indoor mobility \cite{9448014}, and accordingly in \cite{wu2023efficient}, a novel generator to decompose the mobile RIS channel into several components was presented. The generator includes power-law governing user destination selection, behavioral patterns during walking, device random orientation patterns, and fine-grained channel gain calculation via ray-tracing accelerated by the shadow region method. These components, from macro to micro, are separately validated and would jointly characterize the evolution pattern of the mobile RIS channel\footnote{Macro-scale mobility employs a semi-Markov renewal process integrating return regularity and bounded L\'evy-walk. Resident nodes are selected via truncated Pareto distribution with displacement exponent as $0.5$. Sojourn time follows another truncated Pareto distribution with exponent as $1$. Micro-scale trajectories apply steering behavior with four virtual forces: seek, arrival, UE-avoidance, and obstacle-avoidance. UE orientation uses Laplace-distributed polar angles with a mean of $45.11$ and standard deviation of $7.84$ for sitting and Gaussian with a mean of $31.79$ and standard deviation of $7.61$ for walking.}.

Then, we try to find the optimal statistical model (distribution and parameters) that fits the channel. As demonstrated in Fig.~\ref{fig:f2}(b), we enforce a model-driven approach with Nakagami distribution to fit the RIS channel gain statistics at each time step; and in each step, the fit parameters are optimized. Nakagami is the optimal distribution for this model problem though an exhaustive search among the mainstream continuous distributions. In this article, we mainly investigate the AP-RIS and RIS-UE links in the mobile RIS channel. Under far-field conditions, the path reflected by the tile located at the midpoint relative to the AP and UE is equivalent to mirror reflection, similar to the LoS path in traditional channels. Other tiles provide farther propagation distances and larger incidence and reflection angles, forming a response similar to the NLoS path in traditional channels.

Unfortunately, conventional modeling methods have failed when dealing with RIS at scales sufficient to cover buildings. To quantify the inaccuracy, the Kolmogorov-Smirnov distances (KSD) are given along time as $D^{*}=\max _{x}(|\hat{F}(x)-F_\text{N}(x)|),$ where $\hat{F}(x)$ is the empirical cumulative distribution function (CDF) and $F_\text{N}(x)$ is the CDF of the hypothesized Nakagami distribution. Fig.~\ref{fig:f2}(b) shows that even under the optimized fitting, the maximum fitting error goes up to 20\%.  Moreover, different bands lead to different error trends, while different building layouts yield different error evolution. This means in order to use a generalized channel model, for instance, the Nakagami model, we have to train and refine the fitting parameters whenever the setup or time changes. To this end, a data-driven methodology for RIS channel characterization under multi-user mobility is inevitable, which outperforms conventional model-driven approaches in convenience (complexity), accuracy, and flexibility.

The statistical inaccuracy impedes data-driven and learning-based optimization for RIS-covered buildings in terms of concept drift, which unfolds a spatio-temporal perspective. The most obvious one is spatial drift since the building layouts determine the crowd trajectories and then their discrepancy results in the non-uniform distributions. The usually overlooked one hides in temporal, caused by the mobility evolution, as in significant variations of channel statistics over time. It is confirmed in the sub-figure of Jensen-Shannon (JS) divergence in Fig.~\ref{fig:f2}(b), where the higher divergence means more serious drift. This phenomenon not only undermines model-driven methodologies but also poses significant challenges to learning-based approaches, specifically the issue of temporal domain generalization. It calls for a deeper understanding on the spatio-temporal drift in RIS channels.

\begin{figure}
\centering
\includegraphics[width=0.5\textwidth]{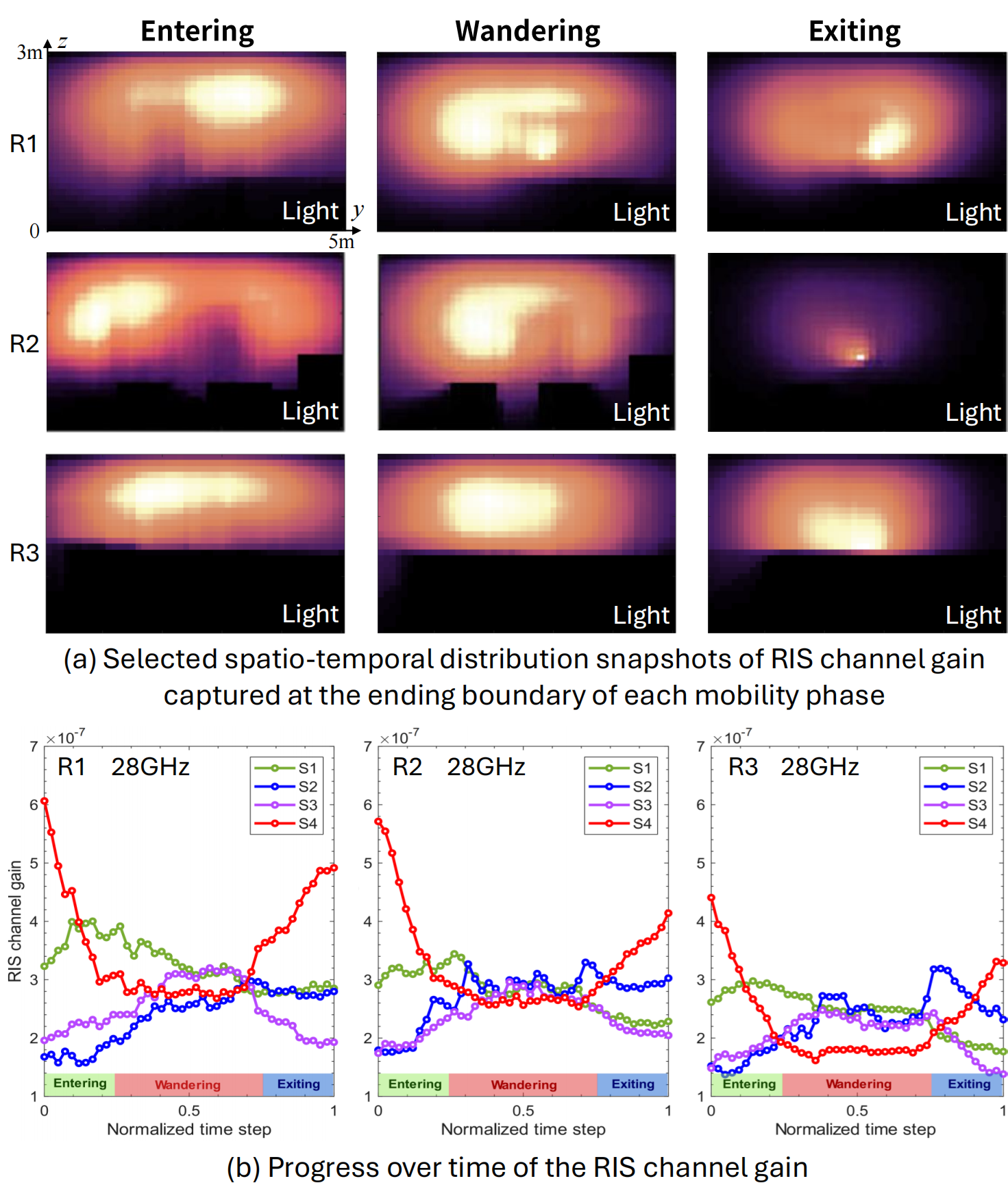}
\caption{(a) Selected spatio-temporal distribution snapshots of RIS channel gain captured at the ending boundary of each mobility phase. The distribution has been normalized in each snapshot, where the brightest region yields the highest channel gain or survival rate, and the darkest one has the lowest value. (b) Progress over time of the RIS channel gain. The illustrated statistics are obtained based on the S4 in the layouts R$1$ to R$3$ supporting $8$ UEs via $4$ APs.}
\label{fig:f4} 
\end{figure}

\section{Emergence of Tidal-like Concept Drifts in RIS Channel Statistics}

Driven by human mobility and constrained by buildings, the statistics of the AP-RIS-UE channels reflect a projection of human mobility in space and time. In this section, we show how the RIS channel distribution on the RIS varies cyclically over time. Like tides, the projection of crowd mobility onto the RIS-covered walls drives the concept drift of surface channel statistics, which is dominated by the macro scale mobility under truncated Pareto distributions \cite{9448014}. Moreover, this tide-like characteristic is reflected in various statistical perspectives of the surface channel, such as the distribution of the outage probability and the probability distribution of the channel gain. 

\subsubsection*{\textbf{Remark}}

The spatio-temporal pattern presented in this section can be found in the statistics for all the layouts and wavebands. Due to space limitations, we randomly chose some of the results as a demonstration to support our conclusions. The analysis and discussion for the presented setup can be generalized to the others unless otherwise stated.

\subsection{Concept drifts due to human mobility behaviors}

To clearly interpret how crowd mobility results in the three-phase spatio-temporal distribution of RIS channel statistics, we select the statistics shown in Fig.~\ref{fig:f3}(a) and (b) as an example. Four APs operating in the VL band are supporting eight UEs, and the sensitivity of VL to obstacles will better reflect the impact of mobility. In Fig.~\ref{fig:f3}(a) and (b), time indices of different traces are unified and then classified into eight intervals. For the sake of clarity, the distribution has been normalized in each snapshot, where the brightest region yields the highest channel gain or survival rate, and the darkest one has the lowest value. Survival probability (or rate) stands for the probability that a link of AP-RIS-UE is not outage. The channel gain thresholds indicating outages are $2\times 10^{-9}$ for VL, $2\times 10^{-8}$ for $73$ GHz mmWave, and $2\times 10^{-7}$ for $28$ GHz mmWave, which depend on the upper bound of NLoS power level. Otherwise, the comparison becomes difficult because the range of values varies greatly from one phase to another as will be depicted in Fig.~\ref{fig:f4}(b). 

Let us now focus on the drift of distribution patterns shown in Fig.~\ref{fig:f3}. The spatio-temporal pattern and its dependence on UE trajectories are exposed by the channel statistics evolution throughout the motion of user's entering, wandering, and exiting the room.  In each phase, the deformation of the bright regions with high-channel gain in each snapshot has a unique pattern. In the entering phase, the shape of the high-channel gain region (top-left) gradually extends from circular to elliptical, and the second high-gain region (top-right) gradually disperses. The transition phase snapshots unveil how the spatio-temporal pattern of RIS channels evolves from the entering to the wandering phase. The shape of such a region remains in the wandering phase, and gradually shrinks with the end of the wandering phase, evolving towards the shape of the region in the existing phase. Meanwhile, on this surface (R$2$-S$4$), we can clearly see the shadow of the interior furnishings. Finally comes the exiting phase, where the shape of the wandering phase region shrinks rapidly, and the high-gain region has shrunk almost to a small point, however, quite unlike the gradual transition from the entering phase to the wandering phase. Such a rapid transition can be further detailed as shown in the snapshots from the transition phase from $70$\% to $72$\% of the complete mobility progress in Fig.~\ref{fig:f3}(a). Irregular vertical grid artifacts appear in the image from the shadow tracks of the users' bodies, which corresponds to the great dynamics of user mobility shifts during the exiting phase, such as frequent UE trajectory crossover and UE avoidance behaviors from each other. The speed of transition between these two phases is so short because the users suddenly turn around and head for the exit. As all users have left, the sub-high gain dispersion area in Fig.~\ref{fig:f3}(a) gradually shrinks. This evolution is like a tidal wave, driven by the evolution of user density distribution.

The channel gain and survival distribution regions in RIS-covered buildings undergo different deformation processes in three phases and vary on different walls due to layout influences, although they both follow the three-phase pattern. As shown in Fig.~\ref{fig:f3}(b), the three-phase evolution is more obvious at the top area in each snapshot. At the entering phase, the area of high survival is small and concentrated in the upper middle. Next, the high survival area spreads out and elongates in parallel directions, as the users' primary area of activity shifts to the indoor central area, and the wandering phase begins. In the wandering phase, the high-survival rate distribution does not seem to change significantly, which is obviously different from the channel gain distribution. In the final exiting phase, the high-survival rate area moves towards the door as users move away towards the exit.

The probability density functions (PDF) variation in Fig.~\ref{fig:f3}(c) shows more clearly the huge dynamic in channel statistics induced by crowd mobility. The channel gain PDF over time has graded widths and erratic peaks. It is such a huge dynamic that leads to the deviation of the traditional fitting method, as aforementioned, which eventually generates a series of non-optimal operational strategies. We also find that even in the same building layout (R$1$), RIS statistics on different walls have shown great differences, since the layouts influence the user's mobility and in turn influence the evolution of RIS channels. We will elaborate on this mechanism in the following subsection.

\subsubsection*{\textbf{Implications}}

This tide-like concept drift will impact the RIS management. For instance, the number of RIS allocations in \cite{wu2024generalized} can be further reduced to only $10$\% in the entering and exiting phases as the available RIS areas in these two phases are far less than in the wandering phase. This is done by a low-complexity deep reinforcement learning-driven optimization.

We found that the impact of human behavior undermines the Markov properties underlying the environment. The first-order Markovian property is a critical dependency for advanced RIS control strategies such as reinforcement learning. As shown in Fig.~\ref{fig:f3}(a), the high partial auto-correlation function (PACF) of channel state with respect to lag reflects the deterioration of the Markovian property. Recent evidence has shown that the tidal evolution caused by human behavior increases the Markovian order of the decision-reward process, which in turn leads to a blow-up in the dimensionality of the state space \cite{wu2024generalized}. Training within a temporally embedded latent decision space, utilizing a solution similar to MuZero \cite{schrittwieser2020mastering} but more lightweight, could potentially generate RIS control strategies that are immune or robust to environmental uncertainties, thereby realizing a truly wireless-friendly building environment.

\subsection{Concept drifts due to various indoor layouts}
\label{sec:waveandenv}

Fig.~\ref{fig:f4}(a) shows the spatio-temporal distributions at the end of each mobility phase under various indoor environments. The most obvious feature is the change in shape of the completely shadowed areas due to the different interior furnishings, which are shown in black. For example, in the R$1$ layout, the conference table is placed in the middle of the room, and there is a certain distance from the wall where the RIS is located, so the edge of the shadow area will be a little fuzzy. In addition, more shadows are distributed in the area far from the door because users enter and leave in the area near the door, resulting in more shadows that UEs induce to the RIS far from the door and under the desks. However, in the layout of R$2$ and R$3$, the interior furnishings are close to the walls, which makes the outline of the completely shadowed area very clear.

Due to the different layout constraints, the impacts of user mobility will also be distinguishable, leading to various evolution patterns of RIS channel statistical distribution. We first focus on the similarities of the evolution patterns. In the three layouts, the area with the highest RIS channel intensity gradually moves down to a lower position over time, and then progressively shrinks, or even disappears. This is mainly because UEs have different orientations in each mobility stage. Specifically, in the entering phase, UE is more likely to point to RIS and the ceiling, and the UE-RIS link is less blocked by the user's body, which raises the high-channel gain area up to a higher position. In the exiting phase, UE is more oriented towards the exit direction, and UE-RIS and RIS-AP links are more easily blocked by the body, which makes the distribution of the high-gain region lower and smaller. In the wandering phase, the two trends mentioned earlier are neutralized, as users perform truncated L\'evy-walk more in the central area.

We investigate the RIS channel gain over time as shown in Fig.~\ref{fig:f4}(b). Starting from the commonality, the average RIS channel gain shares a similar pattern under the three indoor furnishing constraints, i.e., the three-phase pattern that has been emphasized in this work. In the entry and exit phases, the channel gain varies widely and oscillates dramatically. In the wandering phase, however, the RIS channel gain on the different walls fluctuates in a similar interval, as shown in R$1$ and R$2$. Of course, these commonalities essentially stem from the similarity of mobility patterns and interior furnishings. In other words, if the indoor furnishings differ significantly, then the tide-like patterns must also be very different as in R$3$ since R$3$ does not have central gathering areas. In any case, our conclusion remains that there does not exist a static general mathematical model for the RIS channel.

Apart from the large-scale commonality constrained by layout, we now focus on the details hidden in some correlations of layout and mobility patterns in the following aspects.

\begin{itemize}
\item Crossover in S$1$ and S$2$: Note that as the user enters through the door located at S$2$, the UE is facing and close to S$1$, while its back is facing and away from S$2$, so in the entry phase, the RIS channel gain rises first on S$1$ and then becomes smooth, while the RIS channel gain of S$2$ is lower than S$1$. Whereas in the exit phase, the user moves closer to S$2$ and away from S$1$, thus making the RIS channel gain of S$2$ relatively higher than S$1$. Yet during this phase, the AP-RIS-UE link is blocked by users more frequently, so the RIS channel gain eventually decreases.
\item Symmetric in S$3$ and S$4$: Since there is only one door, the user's trajectory is closed. From the perspective perpendicular to S$3$ and S$4$, the user travels from the S$2$ side to the S$1$ side and finally returns to the S$2$ side, which leads to a certain convexity of the curves of S$3$ and S$4$. However, the convexity of curves of S$3$ and S$4$ is clearly symmetric since the S$4$-UE link is blocked with higher probability when the UE is facing S$3$, and vice versa. We find a similarity in Fig.~\ref{fig:f3}(c), since the variation trend of PDF is consistent with Fig.~\ref{fig:f4}(b).
\end{itemize}

\begin{figure}
\centering
\includegraphics[width=0.5\textwidth]{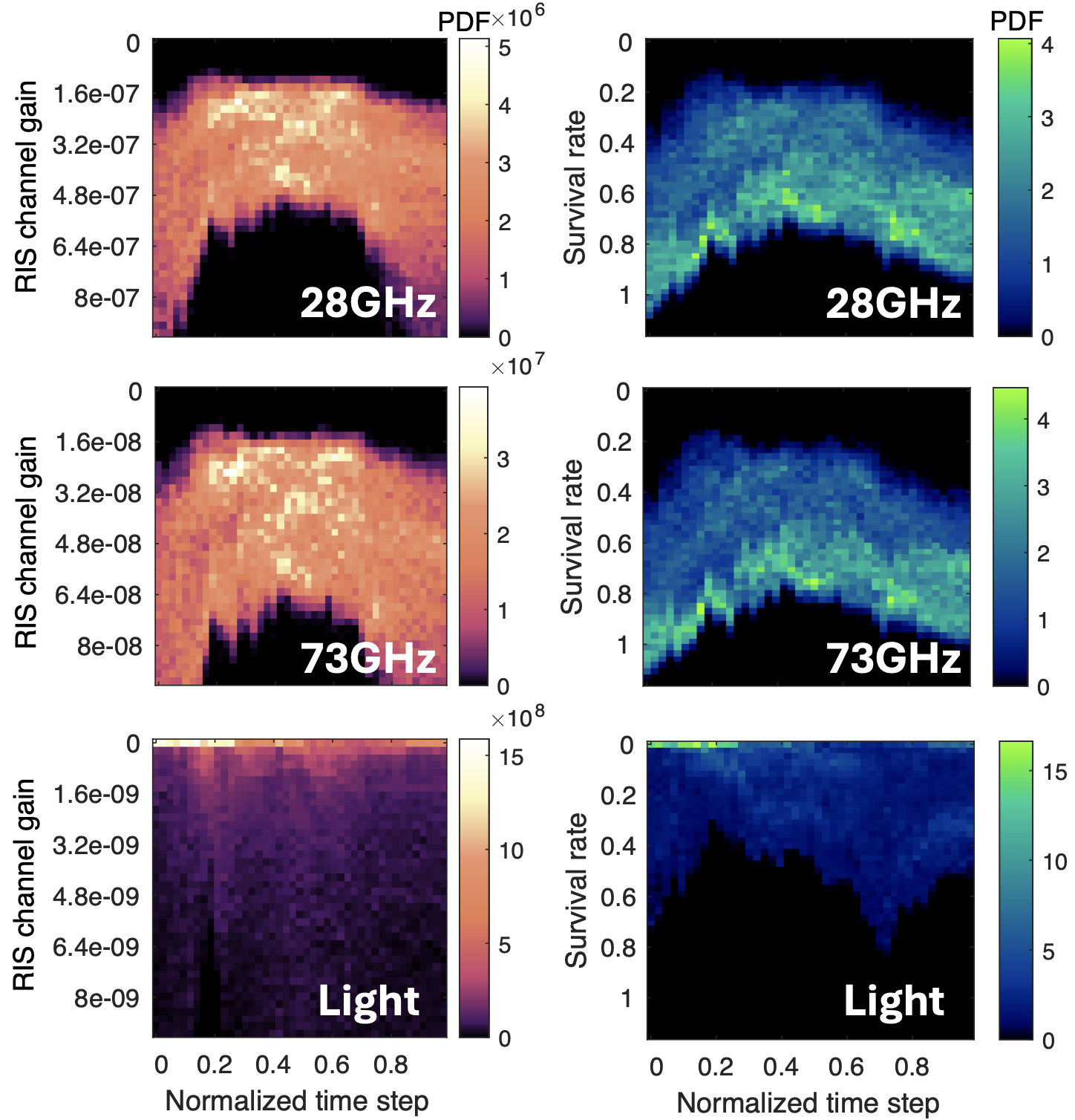}
\caption{Probability density functions of RIS channel gain and survival rate evolving over time under different wavelengths. The demonstrated results are statistically obtained based on the S$4$ in the layouts R$1$ supporting $8$ UEs via $4$ APs. The analysis and discussion for the presented setup can be generalized to the others.}
\label{fig:f5}
\end{figure}

\begin{figure}
\centering
\includegraphics[width=0.5\textwidth]{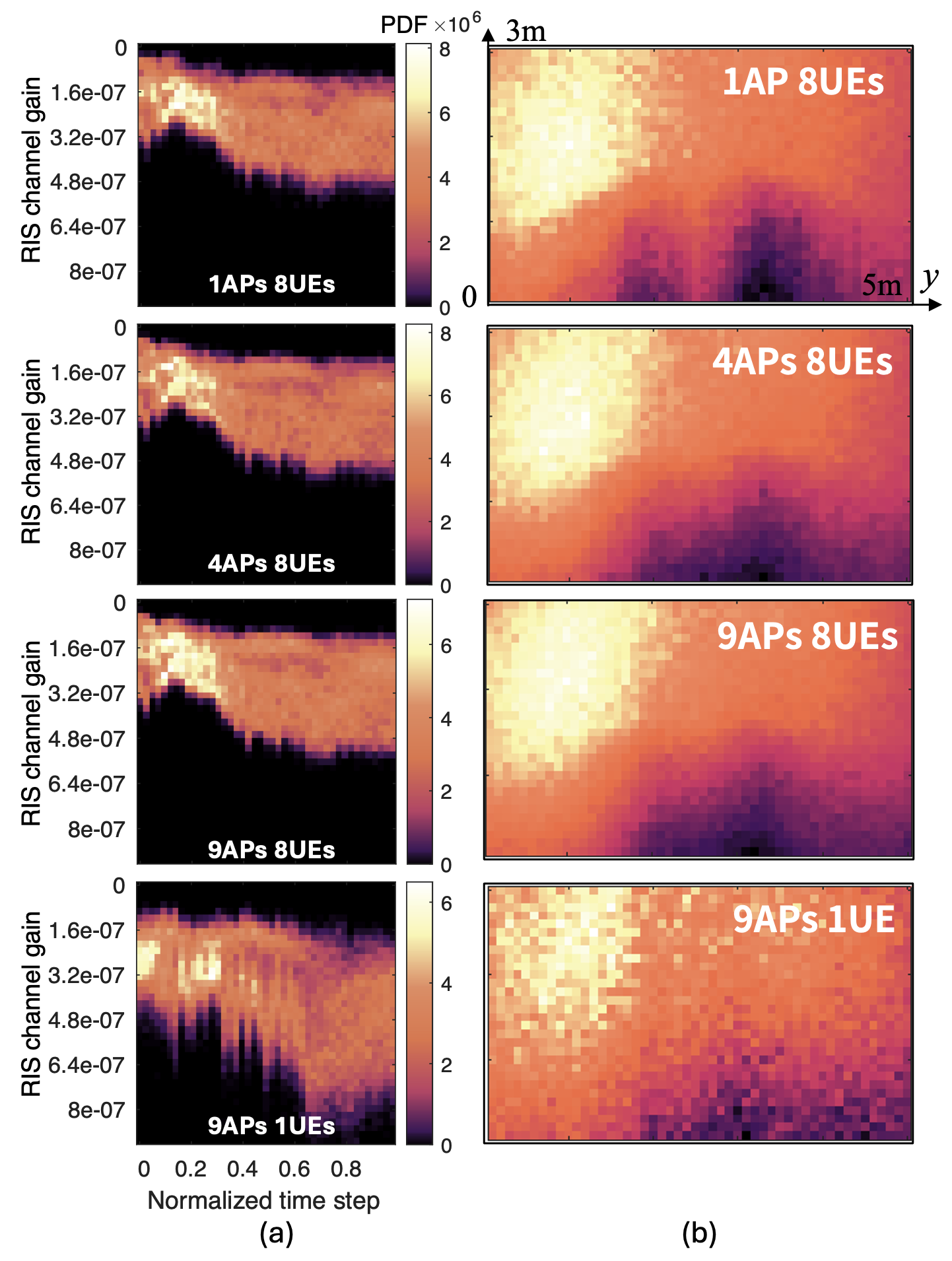}
\caption{(a) The probability density functions of the RIS channel gain and survival rate under different AP to UE ratios. (b) Selected spatial distribution snapshots of RIS channel gain captured at the entering phase with various AP to UE ratios. The illustrated statistics are obtained based on the S$2$ in layout R$1$ operating in the $28$ GHz band. The distribution has been normalized in each snapshot, where the brightest region yields the highest channel gain or survival rate, and the darkest one has the lowest value. The analysis and discussion for the presented setup can be generalized to the others.}
\label{fig:f6}
\end{figure}

\subsubsection*{\textbf{Implications}}

These findings have considerable impacts on the data-driven optimizations of deep learning-based operations. The training of a deep learning method needs to consider each specific layout or even the different walls in the same room. Some layouts or walls have correlated changes, such as opposite or simultaneous trends, so that the neural networks trained in similar environments may be able to generalize to each other. For instance, we adopt the deep point process regressor of channel events in \cite{9448014} to predict AP-RIS-UE channel outages. We train this neural predictor for a central RIS tile on S1 at R1 in $28$ GHz, the valid hit rates are $79.43$\%, $65.32$\%, and $57.90$\% for R$1$ to R$3$. Due to R$2$ being more similar to R$1$ than R$3$, the prediction performance is better in R$2$, however, an adaption of the predictor is needed for R$3$. 

When seeking control decisions, deep reinforcement learning methods are more susceptible to concept drift. Fortunately, recent studies have demonstrated the potential to enhance generalization ability by empowering the agent to perceive concept drift \cite{wu2024concept}. However, there is no free lunch as learning algorithms that generalize to all scenarios simply do not exist. Therefore, efforts to improve generalization should focus on the patterns of human behavior, enabling RIS control strategies that can optimize an entire building wireless environment with minimal training cost.

\subsection{Impacts of wavelength and diffraction ability}

Fig.~\ref{fig:f5} shows us how the RIS channel gain PDF and the survivability PDF evolve for different wavelengths, respectively.
We focus on the effects brought by wavelength and diffraction capability. First, the RIS channel response gradually decreases with increasing frequency, as in Fig.~\ref{fig:f4}(b), which is consistent with our expectation of path loss.  
In mmWave, the increase in frequency causes a significant increase in the channel gain variance. Benefiting from the bypassing capability of mmWave, the effect of outages is more limited. However, in the VL band, the LoS signal is easily blocked, so the majority of the channel gain statistics are distributed in the interval of outages. The remaining component in the PDF of VL bands is very similar to the PDF of NLoS channels in conventional UE-AP links. 
This is because the NLoS links in VL bands are mostly generated from the wall, which exactly corresponds to the RIS tiles.

\subsubsection*{\textbf{Implications}}

Again, we adopt the deep channel event predictor in \cite{9448014} to predict AP-RIS-UE channel outages for a central RIS tile on S$1$. We train this neural predictor at R$1$ in $28$ GHz and then test it using the same model in R1 but for different frequency bands, i.e., $28$ GHz, $73$ GHz, and VL. The valid hit rates are $79.43$\%, $77.14$\%, and $32.62$\% for $28$ GHz, $73$ GHz, and VL. Since there are more frequent outages in VL, and the channel pattern in $73$ GHz is more similar to $28$ GHz, the test performance is better in $73$ GHz.

The intrinsic generalization ability varies significantly across different wavelengths. mmWaves have much better diffraction capabilities compared to VL, resulting in stronger randomness in their channel characteristics. This randomness not only blurs the boundaries of shadowed areas but also obscures the detailed transitions in human behavior patterns. Consequently, algorithms trained for mmWave channels naturally exhibit better generalizations. In contrast, VL channels clearly depict the projection of human behavior in the building environment, inevitably leading to multimodal channel feature evolution and more severe generalization challenges.

\subsection{Impacts of UE to AP ratio}

Usually, we recognize that the increase in AP density will improve signal coverage, but we find some contradictions in the RIS-assisted networks. Fig.~\ref{fig:f6}(b) shows that compared to $1$ AP, the shadowed area increases by about $22\%$ under $4$ APs, and about $27\%$ under $9$ APs. The shadow area increases as the number of APs increases for the following reasons. First, the primary purpose of AP layout is to ensure the coverage of the AP-UE link, so it is common to have evenly distributed APs in the room, such that the shadows cast by APs are misaligned on the same wall. This is because different APs create independent shadows at different areas on RIS, even though some of the shadows may overlap. Therefore, if we cannot jointly optimize the resource allocations for RIS and AP, the overall shadow area on the RIS will increase as the AP number increases. 

\subsubsection*{\textbf{Implications}}

Blindly increasing the number of AP and RIS will reduce the RIS usage rate as the ineffective RIS shadow area is enlarged, but the actual signal quality does not improve significantly, as in Fig.~\ref{fig:f6}(a). This also suggests a new requirement for joint AP-RIS optimization. The preliminary indoor architectural optimization schemes can refer to \cite{zhang2021fundamental}, but the joint optimization across building structure, RIS integration, and AP placement remains a significant challenge.

\section{Conclusions and Open Challenges}

Most of the data traffic takes place within buildings, yet building materials and layouts inherently limit wireless performance. Embedding RIS into building structures that cover the building completely promises fundamental control over BWP. However, this article shows that human behavior introduces complex, evolving channel conditions that resist universal channel models, and thus impede the management tools ranging from conventional optimizations to even deep learning-based methodologies. This study is the first to systematically examine how crowd behavior drives tidal-like shifts in the channels of RIS-covered buildings. 

Our findings show that these dynamics produce high-order Markov dependencies, concept drift, and poor generalization for learning-based prediction and control methods. Therefore, RIS-covered buildings cannot achieve an unlimited increase in BWP, and the upper bound of the performance gains will inevitably be constrained by the tidal evolution patterns revealed herein. This indicates that only the behavioral intelligence of RIS networks can counteract the channel complexity introduced by human behavior and bring us one step closer to the ideal of a wireless-friendly building environment.

For optimizing the RIS-covered buildings under crowd mobility, the viable future solution lies in leveraging reinforcement learning and generative models to organically integrate human behavioral patterns and aesthetic requirements, which are traditionally challenging to quantify, with wireless performance optimization. This approach aims to achieve end-to-end, one-click generation of architectural designs that inherently satisfy predefined functional specifications and performance constraints.

\bibliographystyle{ieeetr}
\bibliography{refs.bib}

\begin{IEEEbiographynophoto}{Zi-Yang Wu} (M'19)
received the B.S. degree in electronic science and technology, the M.S. degree in circuits and systems, and the Ph.D. degree in control science and engineering from Northeastern University, Shenyang, China, in 2014, 2016, and 2020, respectively. He was also a joint Ph.D. student with the Department of Electrical and Computer Engineering, Texas A\&M University, College Station, USA, from 2018 to 2019. He is currently an Associate Researcher with the College of Information Science and Engineering at Northeastern University.
\end{IEEEbiographynophoto}

\begin{IEEEbiographynophoto}{Muhammad Ismail} (S’10-M’13-SM’17)
received the B.Sc. (Hons.) and M.Sc. degrees in electrical engineering (electronics and communications) from Ain Shams University, Cairo, Egypt, in 2007 and 2009, respectively, and the Ph.D. degree in electrical and computer engineering from the University of Waterloo, Waterloo, ON, Canada, in 2013. He is the Director of the Cybersecurity Education, Research, and Outreach Center (CEROC) and an Associate Professor with the Department of Computer Science, Tennessee Technological University, Cookeville, TN, USA.
\end{IEEEbiographynophoto}

\begin{IEEEbiographynophoto}{Jiliang Zhang} (M’15-SM’19) 
is currently a full Professor at College of Information Science and Engineering, Northeastern University, Shenyang, China. He has pioneered systematic building wireless performance evaluation, modeling, and optimization, with the key concepts summarized in ``Fundamental Wireless Performance of a Building'', \textit{IEEE Wireless Communications}, 29(1), 2022.
\end{IEEEbiographynophoto}

\begin{IEEEbiographynophoto}{Jie Zhang}
is the Founder and Chief Scientific Officer of Ranplan Wireless, a publicly listed company on Nasdaq First North, and Cambridge AI+ Ltd, both based in Cambridge, U.K. He has held the Chair in Wireless Systems at the University of Sheffield since 2011. Ranplan Wireless develops Ranplan Professional, one of the world's leading digital twin platforms for wireless environments. Supporting true multi-path ray tracing/launching, Reconfigurable Intelligent Surface (RIS) modelling, extremely large MIMO (xMIMO) and enriched with robust APIs, Ranplan Professional is particularly well-suited for both academic and industrial research into 6G technologies. It also enables the study of low-altitude wireless communications, an emerging domain of critical importance. He been a pioneer in the field of small cell and heterogeneous networks since 2005, and has led ground-breaking work on smart radio environments since 2010.
\end{IEEEbiographynophoto}

\end{document}